\documentclass[12pt,preprint]{aastex}
\usepackage{graphics}
\input epsf
\begin{document}

\title{Discovery of Face-On Counterparts of
Chain Galaxies in the Tadpole ACS Field}

\author{Debra Meloy Elmegreen \affil{Vassar College,
Dept. of Physics \& Astronomy, Box 745, Poughkeepsie, NY 12604;
elmegreen@vassar.edu} }

\author{Bruce G. Elmegreen \affil{IBM Research Division, T.J. Watson
Research Center, P.O. Box 218, Yorktown Heights, NY 10598, USA,
bge@watson.ibm.com} }

\author{Amelia C. Hirst \affil{Vassar College,
Dept. of Physics \& Astronomy, Box 745, Poughkeepsie, NY 12604;
amhirst@vassar.edu} }

\begin{abstract}
The background field of the Tadpole galaxy image taken with the
Hubble Space Telescope ACS contains 87 clump-clusters that are
probably low-inclination versions of the 69 chain galaxies we
found in a previous paper. This conclusion follows from the
similarity in colors and apparent magnitudes of the brightest
clumps in the clump-clusters and chain galaxies, and from the
similarity between the distribution of axial ratios for the
combined sample and the distribution for normal disk galaxies.
These distributions are consistent with chains and clump-clusters
being disks viewed at random angles with intrinsic axial ratios of
$0.1-0.2$ or less. The disks appear to be gas-rich galaxies at z=1
to 2 that are forming an early generation of stars in a small
number of very large star complexes.
\end{abstract}

\keywords{stars: formation ---
galaxies: bulges ---
galaxies: formation ---
galaxies: evolution ---
galaxies: high-redshift ---
galaxies: irregular}

\section{Introduction}

Deep high resolution images of distant galaxy fields, such as the
GOODS fields (Giavalisco et al. 2004) which include the Hubble
Deep Field North (Williams et al. 1996) and the Chandra Deep Field
South (Volonteri, Saracco, \& Chincarini 2000), and the Hubble
Deep Field South (Casertano et al. 2000), provide direct evidence
for processes related to galaxy formation and evolution. A variety
of unusual morphologies is evident, including linear structures
called chain galaxies (Cowie, Hu, \& Songaila 1995; van den Bergh
et al. 1996). In the GOODS-South field, Moustakas et al. (2004)
identified 275 extremely red objects and classified 41 of them as
``others,'' which included chain and low surface brightness
galaxies. The spectral energy distributions of these ``others''
from U through K band did not distinguish between elliptical,
disk, and irregular types even though their visual appearances
varied greatly.

To study chain galaxies in more detail, we surveyed faint linear
features in the Hubble Space Telescope Advanced Camera for Surveys
(HST ACS) field of the Tadpole galaxy, UGC 10214, finding 69 chain
galaxies, 37 double-core galaxies, 21 tadpole-shaped galaxies, and
32 edge-on normal galaxies distinguished by a central bulge and
approximately exponential disk (Elmegreen et al. 2004; hereafter
Paper I). All of the chain, double, and tadpole systems were shown
to have similarly bright clumps, probably the result of star
formation, that are bluer and fainter than the centralized bulges
of most normal galaxies. The chain galaxies were also shown to
dominate all linear types at apparent I-band AB magnitudes fainter
then 24th mag. We suggested that chains are edge-on irregular
galaxies that have no bulges or exponential disks (see also
Dalcanton \& Schectman 1996, Reshetnikov, Dettmar, \& Combes
2003). Two possible examples of faint face-on chains were shown.

We have now surveyed the ACS field of the Tadpole Galaxy for all
faint objects larger than 0.5 arcsec, not only those with linear
structures. These include 87 possible face-on versions of chain
galaxies, 191 normal spiral galaxies of all inclinations, and 34
round, centrally-concentrated galaxies that could be ellipticals.
The face-on versions of chain galaxies appear mostly as small
collections of blue clumps with very faint emission between. To
avoid the presumption that they are complete galaxies before
spectra or deeper images are available, we call them
``clump-clusters," as in Paper I. The normal galaxies are
distinguished by their bright central concentrations of light,
which are presumably bulges, and by their approximately
exponential disks.

We show here that the clumps in clump-clusters have the same
colors and magnitudes as the clumps in linear chain galaxies, that
both types of clumps are distinct from the bulges in normal
galaxies (which are apparently brighter and extend to redder
colors), and that clump-clusters are slightly brighter than chain
galaxies -- presumably as a result of projection -- even though
their average surface brightnesses are comparable or slightly
fainter. We also show that the distribution of axial ratios for
chain galaxies and clump-clusters is similar to the distribution
for local disk galaxies. These observations demonstrate that chain
galaxies are edge-on irregular disks, and their face-on
counterparts are visible as clump-clusters. The properties of the
chain/clump-cluster systems are then reviewed.

\section{Observations}

Archival HST ACS images of UGC 10214 (Tran et al. 2003) were
visually surveyed in the V (F606W) and I (F814W) filters to
examine galaxies in the deep background field.  All objects with a
major axis greater than 10 pixels (0.5 arcsec) were included, 470
total. Axial ratios were determined from contour levels $2 \sigma$
above sky noise. Photometry was done in each of the g (F475W), V,
and I filters using the IRAF task {\it imstat} to define a box
encompassing each galaxy at the $1 \sigma$ contour level, which
corresponds to a surface brightness of 25.5 mag arcsec$^{-2}$.
Conversions to AB magnitudes are given in Paper I. Photometry was
also done on any bright or centralized bulges or hotspots. Radial
profiles were made in I band along the major axis of each galaxy
using the IRAF task {\it pvector} with 3-pixel-wide strips.
Galaxies were divided into the four morphological categories of
Paper I based on the V and I-band images, profiles, and contour
maps. These categories are: chain (70 galaxies), double (49),
tadpole (39), and spiral (191), in addition to elliptical (34) and
clump-cluster (87).

\section{Results}

\begin{figure}
\plotone{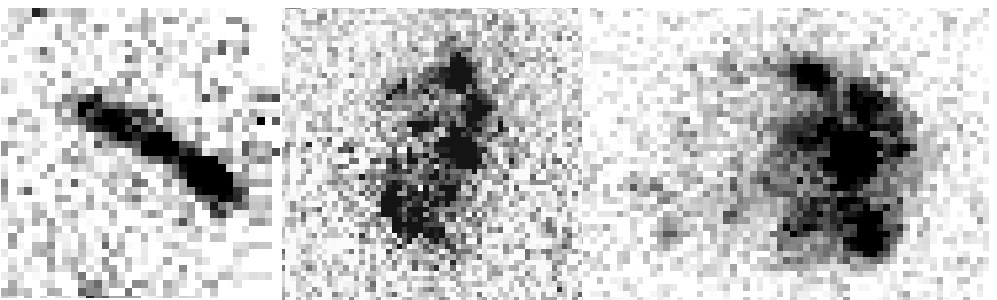} \caption{A chain galaxy (left) and two sample
clump-clusters are shown in I-band from the HST ACS field of the
Tadpole galaxy. The images measure 1.8'', 2.85'' and 3.05''
left-to-right, respectively. }\label{fig:examples}\end{figure}

Figure \ref{fig:examples} shows a sample chain galaxy on the left
and two sample clump-clusters on the right (see also Paper I for
other examples). The pixel sizes are 0.05'' and the fields measure
1.8'', 2.85'', and 3.05'' in the left-right dimension. As is
typical for chain galaxies and clump-clusters, all three systems
are dominated by a small number of very large blue clumps.  We are
proposing here that chain galaxies are edge-on projections of the
clump-clusters, and if this is the case, then the overall
properties of the clumps and whole systems should be similar.

\begin{figure}
\plotone{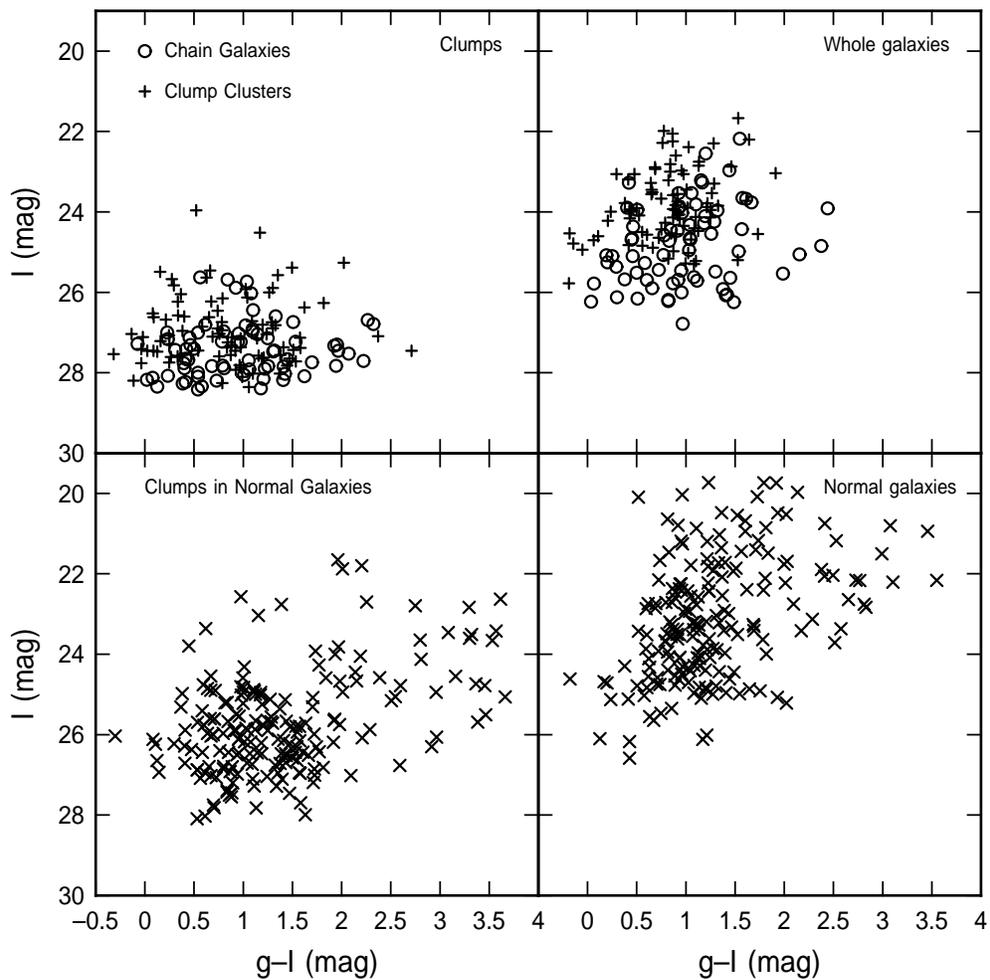} \caption{Color-magnitude diagrams are shown for
the brightest star-forming clumps in the chain galaxies and
clump-clusters (top left) and for the integrated chain and
clump-cluster systems (top right), indicated by different symbols.
CMDs are shown for centralized clumps (i.e., bulges) in normal
galaxies in the deep field (bottom left) and for integrated normal
galaxies (bottom right).}\label{fig:cm}\end{figure}

Figure \ref{fig:cm} shows color-magnitude diagrams (CMD) of the
brightest clumps in chain galaxies and clump-clusters (on the top
left) and of the whole chain and clump-cluster systems (top
right), with different types distinguished by different symbols.
The bottom panels show the CMDs for centralized clumps in normal
galaxies and for the whole normal galaxies in the same deep field.
The clumps in chain galaxies and clump-clusters have the same
distribution of colors and magnitudes, and these colors overlap
the distribution in normal galaxies, which have a prominent red
component in addition. The region of overlap is somewhat blue,
suggesting that clumps in chains, clump-clusters, and normal
galaxies are star-forming regions.  The clumps in chain and
clump-cluster systems average about 1 magnitude fainter than in
the normal galaxies, as do the whole chain and clump-cluster
systems compared to the whole galaxies.

\begin{figure}
\plotone{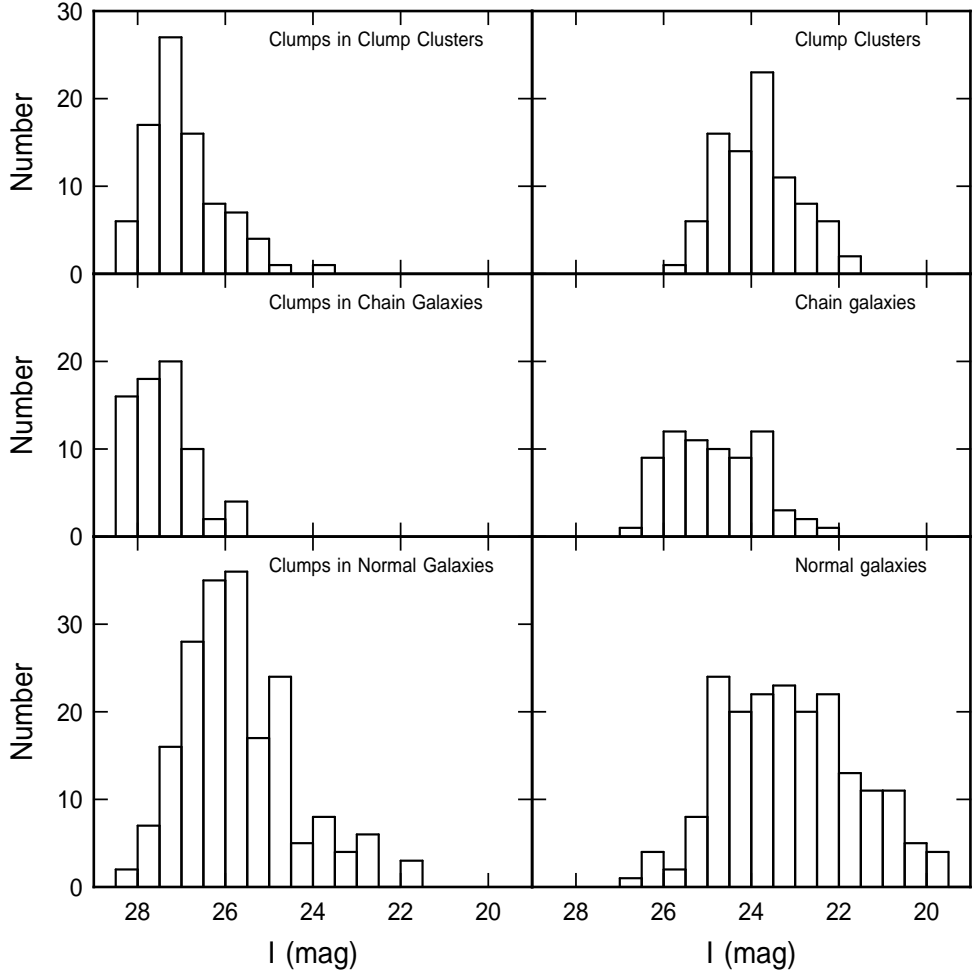} \caption{I-band AB magnitude distributions are
shown for all of the bright or centralized clumps and for whole
galaxies.} \label{fig:hismag}\end{figure}

Figure \ref{fig:hismag} shows the I-band AB magnitude
distributions for all the clumps and whole galaxies.  As suggested
by Figure \ref{fig:cm}, the clumps in the chain and clump-cluster
systems are virtually identical and they are about 1 magnitude
fainter than in normal galaxies. Whole chain galaxies are also
fainter than whole clump-clusters by about 1 magnitude. This is to
be expected if chain galaxies are edge-on projections of
clump-clusters: the larger projected surface areas of the
clump-clusters give them a higher absolute brightness because of
extinction in the edge-on disks, even though clump-clusters can
have slightly fainter average surface brightnesses because of the
smaller line-of-sight path lengths through their disks (Paper I).
In fact, the average clump cluster included here has about the
same surface brightness as the average chain galaxy: 24.0 mag
arcsec$^{-2}$; this is likely to be a reflection of their
dominance near the brightness limit of the survey (which is 25.5 mag
arcsec$^{-2}$ at $1 \sigma$; Paper I). Figure \ref{fig:hismag}
also indicates that whole galaxies are brighter than chains and
clump-clusters (in the right-hand figure) by about the same margin
as for the clumps (in the left-hand figure), suggesting a closer
distance or larger size for normal galaxies.

\begin{figure}
\plotone{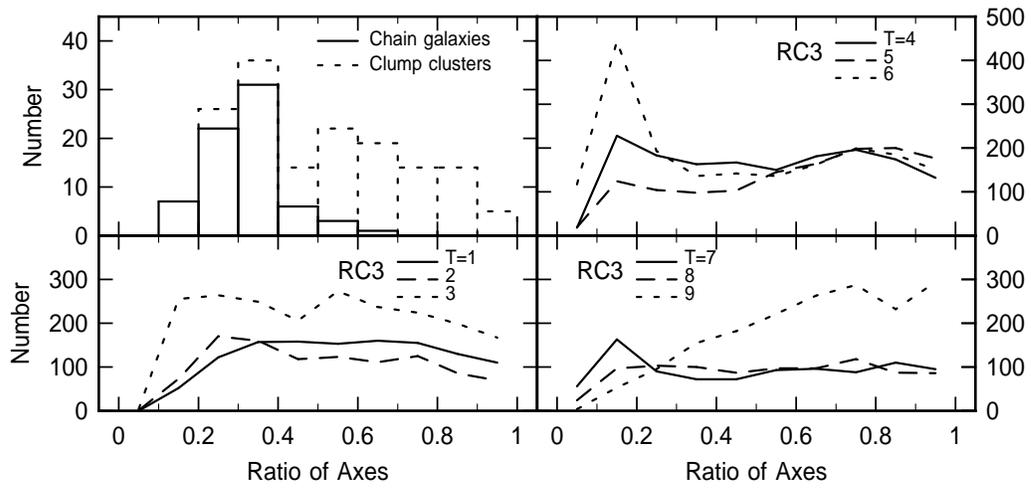} \caption{The distribution of the ratio of axes is
shown for chains and clump-clusters on the top left and for nearby
disk galaxies from the RC3 in the other panels, distinguished by
spiral types T.}\label{fig:hisrat}
\end{figure}

The distribution of the ratio of axes $r$ is shown in Figure
\ref{fig:hisrat}. For a uniform population of thin disks viewed at
random angles, this distribution should be flat between 0 and 1.
For a thick ellipsoidal disk with intrinsic ratio of axes $r_0$,
it should follow the curve
$\left([1-r_0^2][1-r_0^2/r^2]\right)^{-1/2}$ (Sandage, Freeman, \&
Stokes 1970), which is $\sim$constant for $r>>r_0$,
$\rightarrow\infty$ for $r\rightarrow r_0$ and $=0$ for $r<r_0$. A
dispersion in $r_0$ for a sample of galaxies broadens the
singularity at $r=r_0$ into a bump with comparable breadth. The
$r$ distribution for combined chains + clump-clusters is shown on
the top left in Figure \ref{fig:hisrat} and the distribution for
all galaxies in the Third Reference Catalogue of Bright Galaxies
(RC3, de Vaucouleurs et al. 1991) is shown in the rest of the
figure, with different lines for different disk types ($T=1$ for
Sa, $T=9$ for Sm, etc). The combined chain + clump-cluster type
has a distribution of axial ratios that has about the same shape
as the distribution for nearby disk galaxies: somewhat flat at
$r>0.4$, a bump at $r\sim0.2-0.4$, and a fall off at $r<0.2$.
Evidently, the chain galaxies are the mostly edge-on members of a
somewhat uniform population of irregular disk galaxies: the
clump-clusters fill in the near-uniform distribution for more
face-on orientations. For the chain + clump-cluster systems the
minimum axial ratio in our sample is $0.14$, which should be
comparable to the intrinsic relative disk thickness convolved with
the instrument resolution.  For nearby normal galaxies, the
minimum value is $\sim2\times$ smaller for $3\leq T\leq8$. Local
irregulars of type $T=9$ are apparently not disky because their
axial ratio distribution rises for large ratios.

\section{Discussion}

Chain galaxies appear to be edge-on versions of clumpy, disky,
nearly-circular irregular galaxies. They have no prominent bulges
or exponential disks like the normal galaxies in the same deep
field (Paper I), and there are no obvious spiral arms, bars, or
rings as in modern disk galaxies. They contain 3 to 5 enormous
clumps, which appear to be star-forming regions $\sim500$ pc in
size given the blue colors and likely range of cosmological $z$
between 1 and 2. These redshifts are consistent with photometric
and spectroscopic redshifts of chain galaxies with similar angular
sizes measured in the HDF, as noted by comparing our objects with
objects in the online interactive redshift images of the HDF North
(Fernandez-Soto, Lanzetta, \& Yahil 1999) and South (Yahata et al.
2000).  At these redshifts, the I passband corresponds to rest
frame ultraviolet where local late-type galaxies also have
prominent star formation (e.g., Windhorst et al. 2002), although
generally with smaller regions.

The disk-like structure of chains and clump-clusters inferred from
the distribution of axial ratios suggests that most of them did
not form by dissipationless mergers of pre-existing clumps. Such
mergers would make more spheroidal systems. On the other hand,
numerical simulations of young galaxy disks show gravitational
instabilities forming $10^9$ M$_\odot$ clumps like those observed
here; the intermediate stages resemble our clump-clusters and the
clumps eventually merge into bulges (Noguchi 1999). Our
observations are consistent with this picture. The large sizes of
our star-forming regions suggest that gaseous velocity dispersions
are high during disk instabilities, perhaps 20-40\% of the
rotation speeds (the ratio of the velocity dispersion to the
rotation speed is approximately equal to the square root of the
intrinsic ratio of axes). The prominence of clumps rather than
spiral arms suggests further that gaseous densities are much
higher than the critical tidal density from the rotation curve,
i.e., background shear is slow compared to unstable growth. Most
likely, the disks are nearly pure gas when the instabilities begin
(Noguchi 1999). If the observed clumps do not merge into bulges,
then they could form the old stellar disks of today's galaxies.
With high velocity dispersions, these old disk components should
be thick.

We are grateful to J. Blakeslee, H. Ford, and J. Mack for
providing reduced images of the Tadpole field, and to an anonymous
referee for helpful comments. B.G.E. is grateful to NSF for
support under the grant AST-0205097.

\newpage

\end{document}